\documentclass[conference]{IEEEtran}
\usepackage{amsfonts}
\IEEEoverridecommandlockouts

\usepackage{epsfig}
\usepackage{graphicx}
\usepackage{psfig}
\usepackage{subfigure}
\usepackage{epsf}
\usepackage{epstopdf}

\usepackage[cmex10]{amsmath}
\usepackage{booktabs}
\usepackage{fancyhdr}

\usepackage{algorithm}
\usepackage{algorithmic}
\usepackage{epstopdf,cite,color}
\usepackage{amsthm}
\usepackage{amssymb}
\newcommand{\bc}{\begin{center}}
	\newcommand{\ec}{\end{center}}
\newcommand{\be}{\begin{equation}}
\newcommand{\ee}{\end{equation}}
\newcommand{\bea}{\begin{eqnarray}}
\newcommand{\eea}{\end{eqnarray}}

\usepackage[letterpaper, left=0.625in, right=0.625in, bottom=1in, top=0.70in]{geometry}

\renewcommand\IEEEkeywordsname{Index Terms}

\begin{document}
	
	\title{Secure and Energy-Efficient Offloading and Resource Allocation in a NOMA-Based MEC Network}
	\author{Qun~Wang\textsuperscript{\dag}, Han Hu$^*$, Haijian Sun\textsuperscript{\ddag}, Rose Qingyang Hu\textsuperscript{\dag}\\ 
		\textsuperscript{\dag}Department of Electrical and Computer Engineering, Utah State University, Logan, UT, USA\\%
		$^*$Jiangsu Key Laboratory of Wireless Communications, \\ Nanjing University of Posts and Telecommunications, Nanjing, China\\
		\textsuperscript{\ddag}Department of Computer Science, University of Wisconsin-Whitewater, Whitewater, WI, USA\\
		Emails: claudqunwang@ieee.org, han\_h@njupt.edu.cn, h.j.sun@ieee.org, rose.hu@usu.edu
		\thanks{The project has been sponsored by the National Science Foundation under grant NSF EARS-1547312 and CNS-2007995.}
	}
	\maketitle
	
	\IEEEpeerreviewmaketitle
\begin{abstract}
Energy efficiency and security are two critical issues for mobile edge computing (MEC) networks. With stochastic task arrivals, time-varying dynamic environment, and passive existing attackers, it is very challenging to offload computation tasks securely and efficiently.
In this paper, we study the task offloading and resource allocation problem in a non-orthogonal multiple access (NOMA) assisted MEC network with security and energy efficiency considerations. To tackle the problem, a dynamic secure task offloading and resource allocation algorithm is proposed based on Lyapunov optimization theory.
A stochastic non-convex problem is formulated to jointly optimize the local-CPU frequency and transmit power, aiming at maximizing the network energy efficiency, which is defined as the ratio of the long-term average secure rate to the long-term average power consumption of all users. The formulated problem is decomposed into the deterministic sub-problems in each time slot. The optimal local CPU-cycle and the transmit power of each user can be given in the closed-from.
Simulation results evaluate the impacts of different parameters on the efficiency metrics and demonstrate that the proposed method can achieve better performance compared with other benchmark methods in terms of energy efficiency.
\end{abstract}
\IEEEpeerreviewmaketitle
\section{Introduction}
The explosive data traffic growth, fast development, and commercialization of the 5G wireless communication networks impose great challenges on data security as well as global energy consumption \cite{5Ghu}. 
In order to improve energy efficiency {(EE)}, mobile edge computing (MEC) and non-orthogonal multiple access (NOMA) have been envisaged as two promising technologies in 5G and the forthcoming 6G wireless networks. By deploying edge servers with high computational capacities close to end users, the end users can offload partial or all computation tasks to the nearby MECs to save power as well as speed up the computing \cite{eemc}. Meanwhile, by exploiting superposition coding at the transmitter and successive interference cancellation (SIC) at the receiver, NOMA brings significant changes to the multiple access. NOMA allows multiple users to share the same radio bandwidth in either power domain or code domain to improve spectral efficiency with a relatively higher receiver complexity \cite{noma}.

Applying NOMA into MEC-enabled networks has recently received extensive attention due to its performance gain in both spectrum efficiency and EE\cite{nec1}-\cite{nec3}. Most of the existing works didn't taking the security issue into account.
In fact, due to the broadcast nature of the wireless link, it could be very vulnerable for the tasks to be intercepted by the eavesdroppers. The physical layer security (PLS) in the NOMA-assisted MEC networks has received many research interests \cite{sechj}. 
The joint consideration of PLS in the NOMA assisted MEC network was studied in \cite{wu}-\cite{secac}.
In \cite{wu}, an iterative algorithm was proposed to maximize the minimum anti-eavesdropping ability in a MEC network with uplink NOMA.
The authors in \cite{wang} proposed a bisection searching algorithm to minimize the maximum task completion time subject to the worst-case secrecy rate. Instead of only considering the power consumption or computing rate performance above, \cite{secac} studied the EE maximization problem for a NOMA enabled MEC network with eavesdroppers.

Most of the existing works on NOMA-assisted MEC with external eavesdroppers typically focus on the performance evaluation in the  scenarios where either channel conditions or required tasks remain constant. Such an assumption makes the analysis on the computation offloading and resource allocation more tractable. However, in a dynamic environment, the dynamic behaviors of the workload arrivals and fading channels impact the overall system performance. Thus the system design that focuses on the short term performance may not work well from the long term perspective. Towards that, a stochastic task offloading model and resource allocation strategy should be adopted \cite{Nouri}. In this paper, we integrate PLS and study the long-term EE performance in a NOMA-enabled MEC network. By incorporating the statistical behaviors of the channel states and task arrivals, we formulate a stochastic optimization problem to maximize the long-term average EE subject to multiple constraints including task queue stability, maximum available power, and peak CPU-cycle frequency. An energy-efficient offloading and resource allocation method based on Lyapunov optimization is proposed. 
The simulation results validate the superior performance of the proposed method in terms of EE in a secure NOMA-assisted MEC network.

The rest of the paper is organized as follows. Section II describes the system model. In Section III, the EE maximization problem and corresponding alternative solution are presented. Numerical results are provided in Section IV. The paper is concluded in Section V.  
\section{System Model}
\begin{figure}[h]
	\vspace{-0.3cm}
	\setlength{\abovecaptionskip}{-0.2cm} 
	\setlength{\belowcaptionskip}{-1cm}
	\centering
	\includegraphics[width=3.0in]{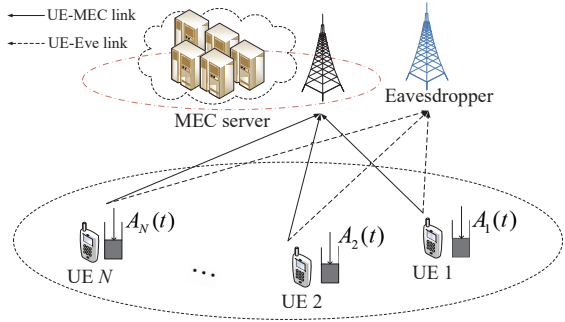}	
	\caption{System Model.\label{symodel}}
\end{figure}
In Fig. \ref{symodel}, an uplink NOMA communication system is considered, which consists of $N$ user equipments (UEs), one access point (AP) with the MEC server, and one external eavesdropper (Eve) near the AP. All the UEs can offload their computation tasks to the MEC while the external eavesdropper intends to intercept the  confidential information. The arrival task of user $n$ at time slot $t$ is denoted as $A_n(t)$. Note that the prior statistical information of $A_n(t)$ is not required and it could be difficult to obtain in the practical systems. We focus on a data-partition-oriented computation task model. A partial offloading scheme is used, i.e., part of the task is processed locally and the remaining part of the data can be offloaded to the remote server for processing. For each UE, local computing and task offloading can be executed simultaneously.

Assuming that each UE has buffering ability, where the arrived but not yet processed data can be queued for the next time slot. Let $Q_n(t)$ be the queue backlog of UE $n$, and its evolution equation can be expressed as
\be
\setlength{\abovedisplayskip}{3pt}
\setlength{\belowdisplayskip}{3pt}
{Q}_n(t + 1) = \max \{ {Q}_n(t) - {R_n^{tot}(t)\tau},0\} + {A}_n(t),
\ee
where $R_n^{tot}(t)=R_n^{off}(t)+R_n^{loc}(t)$ is the total computing rate of UE $n$ at time slot $t$, $R_n^{off}(t)$ and $R_n^{loc}(t)$ are secure offloading rate and local task processing rate, respectively. $\tau$ is time duration of each slot.
\subsection{Local Computing Model}
Let $f_n(t)$ denote the local CPU-cycle frequency of UE $n$, which cannot exceed its maximum value $f_{\max}$. Let $C_n$ be the computation intensity (in CPU cycles per bit). Thus, the local task processing rate can be expressed as $R_n^{loc}(t)= f_n(t)/C_n$. We use the widely adopted model $P_n^{loc}(t)=\kappa_{n}f^3_n(t)$ to calculate the local computing power consumption of UE $n$, where $\kappa_n$ is the energy coefficient and its value depends on the chip architecture \cite{qiot}.

\subsection{Task Offloading Model}
The independent and identically distributed (i.i.d) frequency-flat block fading channel model is adopted, i.e., the channel remains static within each time slot but varies across different time slots. The small-scale fading coefficients from UE $n$ to the MEC server and to the Eve are denoted as ${H_{b,n}}(t)$ and ${H_{e,n}}(t)$, respectively. Both are assumed to be exponential distributed with unit mean \cite{Ymao}. Thus, the channel power gain from UE $n$ to the MEC is given as ${h_{i,n}}(t) = {H_{i,n}}(t){g_0}({d_0}/{d_{i,n}})^\theta$, $i \in \{b,e\}$, where $g_0$ is the path-loss constant, $\theta$ is the path-loss exponent, $d_0$ is the reference distance, and ${d_{i,n}}$ is the distance from UE $n$ to receiver.  Furthermore, to improve the spectrum efficiency, NOMA is applied on the uplink access for offloading. We assume that $h_{b,1}\le h_{b,2}\le \cdots \le h_{b,N}$ and $h_{e,1}\le h_{e,2}\le \cdots \le h_{e,N}$. Using SIC at the receiver side, the achievable secure offloading rate at UE $n$ can be given by
  \be
  \setlength{\abovedisplayskip}{3pt}
  \setlength{\belowdisplayskip}{3pt}
 R_{n}^{off}(t)=[B\log_2(1+\gamma_{b,n})-B\log_2(1+\gamma_{e,n})]^+, 
 \ee
 where $B$ is the bandwidth allocated to each UE, $\gamma_{b,n}=\frac{p_{n(t)}h_{b,n}(t)}{\sum_{i=1}^{n-1}p_i(t)h_{b,i}(t)+\sigma_{b,n}^2}$ and $\gamma_{e,n}=\frac{p_{n}(t)h_{e,n}(t)}{\sum_{i=1}^{n-1}p_ih_{e,i}(t)+\sigma_{e,n}^2}$ are the SINRs received by the MEC server and the Eve respectively. $p_{n}(t)$ is the transmit power of UE $n$, $\sigma_{b,n} ^{2}$ and $\sigma_{e,n} ^{2}$ are the background noise variances at the MEC and the Eve respectively. $[x]^+$= $\max(x,0)$.
 The power consumption for offloading can be expressed as
$ P_n^o(t) = \zeta p_n(t)+p_r$,
where $\zeta$ is the amplifier coefficient and $p_r$ is the constant circuit power consumption.

\section{Dynamic Task Offloading and Resource Allocation}
\subsection{Problem Formulation}
EE is defined as the ratio of the number of long term total computed bits achieved by all the UEs to the total energy  consumption \cite{EEDf},
\be
\setlength{\abovedisplayskip}{3pt}
\setlength{\belowdisplayskip}{3pt}
\eta(t)=\frac{{\lim }_{T \to \infty } \frac{1}{T}\mathbb{E}[\sum_{t=1}^{T}R_{tot}(t)\tau]}{{\lim }_{T \to \infty } \frac{1}{T}\mathbb{E}[\sum_{t=1}^T P_{tot}(t)\tau]}=\frac{\overline{R}_{tot}\tau}{\overline{P}_{tot}\tau},
\ee
where $R_{tot}(t)=\sum_{n=1}^NR_n^{tot}(t)$ and $P_{tot}(t)=\sum_{n=1}^N P_n^{off}(t)+P_n^{loc}(t)$ are the total achievable rate and consumed power by all the users at $t$.

This work aims to maximize the long-term average EE for all the UEs under the constraints of resource limitations while guaranteeing the average queuing length stability. Therefore, the problem is formulated as
\begin{subequations}
	\setlength{\abovedisplayskip}{3pt}
	\setlength{\belowdisplayskip}{3pt}
	\label{P0}
\begin{alignat}{5}
\textbf{P}{_0:}~ &\mathop {\max}_{f_n(t),p_n(t)} \mathop \eta\nonumber\\	 
s.t.~&~~ P_n^{tot}(t)\le P_{\max},\\
& \mathop {\lim }\limits_{T \to \infty } \frac{1}{T}\mathbb{E}[ |\ \overline Q_n(t)| ]=0,\\ 
&{f_n}(t) \le {f_{\max }},\\ 
&0 \le p_n(t),
\end{alignat}
\end{subequations}
where $\overline Q_n(t)$ is the average queue length of UE $n$. The constraint (\ref{P0}a) indicates that the total power consumed by UE at time slot $t$ should not exceed the maximum allowable power $P_{\max}$. (\ref{P0}b) requires the task buffers to be mean rate stable, which also ensures that all the arrived computation tasks can be processed within a finite delay. (\ref{P0}c) is the range of local computing frequency, and (\ref{P0}d) denotes the transmit power of each UE should not be negative.

\subsection{Problem Transformation Using Lyapunov Optimization}
The problem $\textbf{P}_0$ is a non-convex problem, which is difficult to be solved due the fractional structure of the objective function and the long term queue constraint (\ref{P0}b). By introducing a new parameter $\eta^*(t)=\frac{\sum_{i=0}^{t-1}R_{tot}(i)\tau}{\sum_{i=0}^{t-1}P_{tot}(i)\tau}$ \cite{EEDf}, the problem can be transformed to $\textbf{P}_1$, which can be solved in an alternating way.
\begin{subequations}
	\setlength{\abovedisplayskip}{3pt}
	\setlength{\belowdisplayskip}{3pt}
	\label{P1}
	\begin{alignat}{5}
	\textbf{P}{_1:}~  {\max}_{f_n(t),p_n(t)}&  \overline{R}_{tot}(t)\tau- \eta^*(t)\overline{P}_{tot}(t)\tau\nonumber\\	
	s.t.~~&(\ref{P0}a)-(\ref{P0}d).\nonumber
    \end{alignat}
\end{subequations}
{
Note that $\eta^*(t)$ is a parameter that depends on the resource allocation strategy before $t$-th time block \cite{EEDf}.} In the following, the Lyapunov optimization is introduced to tackle the task queue stability constraint.

To stabilize the task queues, the quadratic Lyapunov function is first defined as $L(\mathbf{Q}(t))\mathop  = \limits^\Delta  \frac{1}{2}\sum_{n=1}^N Q_n^2{(t)}$ \cite{lypu}.
Next, the one-step conditional Lyapunov drift function is introduced to push the quadratic Lyapunov function towards a bounded level.
\be
	\setlength{\abovedisplayskip}{3pt}
	\setlength{\belowdisplayskip}{3pt}
\Delta (\mathbf{Q} (t))\mathop  = \limits^\Delta  \mathbb{E}[L(\mathbf{Q} (t + 1)) - L(\mathbf{Q} (t))|\mathbf{Q} (t)].
\ee
By incorporating queue stability, the Lyapunov drift-plus-penalty function is defined as
\be
	\setlength{\abovedisplayskip}{3pt}
	\setlength{\belowdisplayskip}{3pt}
{\Delta _V}(\mathbf{Q} (t)) =- \Delta (\mathbf{Q} (t)) + V [R_{tot}(t)\tau- \eta^*(t)P_{tot}(t)\tau],
\ee
where $V $ is a control parameter to control the tradeoff between the queue length and system EE. The minus sign is used to maximize EE and to minimize the queue length bound.
For an arbitrary feasible resource allocation decision that is applicable in all the  time slots, the drift-plus-penalty function ${\Delta _V}(\mathbf{Q}(t))$ satisfies
\be
\setlength{\abovedisplayskip}{3pt}
\setlength{\belowdisplayskip}{3pt}
\begin{aligned}
	\Delta_V(\mathbf{Q}(t))& \ge -C+\sum_{n=1}^N \mathbb{E}\{Q_n(t)(R_n^{tot}(t)\tau-A_{n}(t))\}\\
	&+V\sum_{n = 1}^N[R_n^{tot}(t)\tau- \eta^*(t)P_n^{tot}(t)\tau],
\end{aligned} 
\ee
where $C = \frac{1}{2}\sum\limits_{u = 1}^U {({R_{n}^{\max }}^2\tau^2 + {A_{n}^{\max}}^2)} $, $R_{n}^{\max }$ and $A_{n}^{\max}$ are the maximum achievable computing rate and the maximum arrival workload, respectively. 

Thus, $\mathbf{P}_1$ is converted to a series of per-time-slot deterministic optimization problem $\mathbf{P}_2$, which needs to be solved at each time slot and is given as in \textbf{Algorithm 1}.
%
%
\begin{algorithm}[!t]
	\algsetup{linenosize=\small}
	\small
	\caption{ Dynamic Resource Allocation Algorithm }
	\label{alg1}
	\begin{algorithmic}[1]
		\STATE At the beginning of the $t$th time slot, obtain $\{Q_n(t)\}$, $\{A_n(t)\}$.
		\STATE Determine $\mathbf{f}(t)$ and $\mathbf{p}(t)$ by solving 
		\be
		\setlength{\abovedisplayskip}{3pt}
		\setlength{\belowdisplayskip}{3pt}
		\begin{aligned}
			\textbf{P}{_2:}~ & \max_{f_n(t),p_n(t)} \sum_{n=1}^N\{Q_n(t)(R_n^{tot}(t)\tau-A_{n}(t))\}\\
			&+V\sum_{n = 1}^N[R_n^{tot}(t)\tau- \eta^*(t)P_n^{tot}(t)\tau]\\
			s.t.~~& (\ref{P0}a),(\ref{P0}c),(\ref{P0}d) \nonumber\\
		\end{aligned}
		\ee
		\STATE Update $\{Q_n(t)\}$ and set $t=t+1$. Go back to step 1. 
	\end{algorithmic}
\end{algorithm}
In $\textbf{P}_{2}$, $\mathbf{f}(t)$ and $\mathbf{p}(t)$ can be decoupled with each other in both the objective function and the constraints. Thus, the problem $\textbf{P}_{2}$ can be decomposed into two sub-problems, namely the optimal CPU-cycle frequency scheduling sub-problem and the optimal transmit power allocation sub-problem, which can be solved alternately in the following. 

\textbf{Optimal CPU-Cycle Frequencies Scheduling:}
The optimal CPU-cycle frequencies  $\mathbf{f}(t)$ can be obtained by
\begin{alignat}{5}\label{PC}
	\setlength{\abovedisplayskip}{3pt}
	\setlength{\belowdisplayskip}{3pt}
\textbf{P}_{2.1}{:}\max_{ 0\le f_n(t)\le f_{\max}}& \sum_{n=1}^N(Q_n(t)+V)(R_n^{off}(t)+f_n(t)/C_n)\nonumber\\
&- V\eta^*(t)(\kappa_{n}f_n^3(t)+p_r+\zeta p_n(t))\nonumber\\
s.t.~~&\kappa_n f_n^3(t)\le P_{\max}-P_n^{off}.
\end{alignat}
Since the objective function of $\mathbf{P}_{2.1}$ and the constraints are convex with respect to $f_n(t)$, the optimal $f_n(t)$ can be given as 
\be
	\setlength{\abovedisplayskip}{3pt}
	\setlength{\belowdisplayskip}{3pt}
f_n^* = \left[\sqrt{\frac{(V + {Q_n}(t))}{3V\eta \kappa_n {C_n}}} \right]_0^{\overline{f}_{\max}},
\ee
where $\overline{f}_{\max}=\min\{{f_{\max}},\root 3 \of {(P_n^{\max} - \zeta {p_n} - {p_r})/ \kappa_n}\}$ is the upper bound of the frequency.

\textbf{Optimal Transmit Power Allocation:}
For the transmission power allocation optimization, the problem $\mathbf{P}_2$ is transformed into
	\begin{alignat}{5}\label{PW3}
		\setlength{\abovedisplayskip}{3pt}
	\setlength{\belowdisplayskip}{3pt}
	\textbf{P}_{2.2}{:}~  \max_{{p}_n(t)}&\sum_{n=1}^NB\ln 2(Q_n(t)+V)[\ln({ \sum \limits_{i = 1}^{n} {p_i}(t)h_{b,i}^2 + \sigma _{b,n}^2})\nonumber\\
	&-\ln({ \sum \limits_{i = 1}^{n - 1} {p_i}(t)h_{b,i}^2 + \sigma _{b,n}^2})- \ln( { \sum \limits_{i = 1}^{n} {p_i}h_{e,i}^2 + \sigma _{e,n}^2})\nonumber\\
	&+\ln({ \sum \limits_{i = 1}^{n - 1} {p_i}h_{e,i}^2 + \sigma _{e,n}^2})+\frac{f_n}{B\ln 2C_n}]\nonumber\\
	&- V\eta^*(t)(\zeta {p_n} + {p_r} + \kappa_n {f^3_n})\nonumber\\	
	s.t.~~&0\le {p_n}(t) \le (P_{\max}-{p_r} - \kappa_n {f^3_n})/\zeta.
	\end{alignat}
The minus logarithmic terms make the objective function not convex, which is addressed by  Lemma 1 introduced in the following.

$\mathbf{Lemma~1}$: By introducing the function $\phi(y)=-yx+\ln y+1$,  $\forall x > 0$, one has
\be
	\setlength{\abovedisplayskip}{3pt}
	\setlength{\belowdisplayskip}{3pt}
-\ln x=\max_{y>0}\phi(y).
\ee
The optimal solution can be achieved at $y=1/x$. The upper bound can be given by using Lemma 1 as $\phi(y)$ \cite{lemma1}. By setting $y_{b,n}={ \sum \limits_{i = 1}^{n - 1} {p_i}(t)h_{b,i}^2 + \sigma _{b,n}^2}$ and $y_{e,n}={ \sum \limits_{i = 1}^{n} {p_i}(t)h_{e,i}^2 + \sigma _{e,n}^2}$, one has 
	\begin{alignat}{5}\label{PW4}
		\setlength{\abovedisplayskip}{3pt}
	\setlength{\belowdisplayskip}{3pt}
	\textbf{P}_{2.3}{:}~  &\max_{{p}_n(t),y_{b,n},y_{e,n}}\sum_{n=1}^NB\ln 2(Q_n(t)+V)[\ln( \sum \limits_{i = 1}^{n} {p_i}(t)h_{b,i}^2\nonumber\\
	&+ \sigma_{b,n}^2)+\phi_{b,n}(y_{b,n})+\phi_{e,n}(y_{e,n})+\ln({ \sum \limits_{i = 1}^{n - 1} {p_i}(t)h_{e,i}^2 + \sigma _{e,n}^2})\nonumber\\
	&+\frac{f_n}{B\ln 2C_n}]- V\eta^*(t)(\zeta {p_n}(t) + {p_r} + \kappa_n {f^3_n})-Q_n(t)A_{n}(t)\nonumber\\	
	&s.t.~~0\le {p_n}(t) \le (P_{\max}-{p_r} - \kappa_n {f^3_n})/\zeta,
	\end{alignat}
where $\phi_{b,n}(y_{b,n})=-y_{b,n}({ \sum \limits_{i = 1}^{n - 1} {p_i}(t)h_{b,i}^2 + \sigma _{b,n}^2})+\ln y_{b,n}+1$, and $\phi_{e,n}(y_{e,n})=-y_{e,n}({ \sum \limits_{i = 1}^{n} {p_i}(t)h_{e,i}^2 + \sigma _{e,n}^2})+\ln y_{e,n}+1$.
The problem $\textbf{P}_{2.3}$ is a convex problem with respect to both $p_n(t)$ and $y_{b,n},y_{e,n}$. It can be solved by using a standard convex optimization tool. After we obtain $p_n^*(t)$, the values of $y_{b,n}^*$ and $y_{e,n}^*$ can be respectively given by $y_{b,n}^*=({ \sum \limits_{i = 1}^{n - 1} {p_i^*}(t)h_{b,i}^2 + \sigma _{b,n}^2})^{-1}$ and $ y_{e,n}^*=({ \sum \limits_{i = 1}^{n} {p_i^*}(t)h_{e,i}^2 + \sigma _{e,n}^2})^{-1}$.
By alternately updating $p_n(t)$ and $y_{b,n},y_{e,n}$, the optimal solutions of $\mathbf{P}_{2.3}$ can be achieved at convergence.

\textbf{Remark 1:} To obtain fundamental and insightful understanding of the offloading power allocation for a multi-user NOMA assisted secure MEC system, we consider a special case with two UEs \cite{fzsec}. The problem with respect to  $p_n$ is given as
	\begin{alignat}{5}\label{PW5}
		\setlength{\abovedisplayskip}{3pt}
	\setlength{\belowdisplayskip}{3pt}
		\textbf{P}_{2.4}{:}~  &\max_{p_1(t),p_2(t)}B\ln 2(V + {Q_2}(t))[\ln ({p_2}(t)h_{b,2}^2 + {p_1}(t)h_{b,1}^2 + \sigma _{b,2}^2)\nonumber \\
	&- \ln ({p_1}(t)h_{b,1}^2 + \sigma _{b,2}^2)- {y_{b2}}({p_1}(t)h_{b,1}^2 + \sigma _{b,2}^2)\nonumber\\
	&+ \ln {y_{b2}} + 1 + \ln ({p_1}(t)h_{e,1}^2 + \sigma _{e,2}^2) + {{{f_2}} \over {{C_2}B\ln 2}}]\nonumber\\
	 &+ B\ln 2(V+ {Q_1}(t))[\ln (\sigma _{b,1}^2 + {p_1}(t)h_{b,1}^2) - \ln \sigma _{b,1}^2\nonumber\\
	 &- {y_{e1}}(\sigma _{e,1}^2 + {p_1}(t)h_{e,1}^2) + \ln {y_{e1}} + 1+\ln \sigma _{e,1}^2\nonumber\\
	 & + {{{f_1}} \over {{C_1}B\ln 2}}] - V\eta (\zeta ({p_2}(t) + {p_1}(t)) + 2{p_r} + \kappa_n ({f_n}^3)\nonumber\\
		s.t.~~&0\le {p_n}(t) \le (P_{max}-{p_r} - \kappa_n {f_n^3})/\zeta.
	\end{alignat}
$\textbf{P}_{2.4}$ is a convex problem with respect to $p_1(t)$ and $p_2(t)$, and the optimal solutions are given as
\be
	\setlength{\abovedisplayskip}{3pt}
	\setlength{\belowdisplayskip}{3pt}
{p^*_1}(t) = {{ - b_1 \pm \sqrt {b_1^2 - 4{b_2}} } \over 2},
\ee
and 
\be
	\setlength{\abovedisplayskip}{3pt}
	\setlength{\belowdisplayskip}{3pt}
{p^*_2(t)}{\rm{ = }}{1 \over {({{V\eta \zeta } \over {B\ln 2(V + {Q_2}(t))}} + {y_{e2}}h_{e,2}^2)}} - \frac{{p_1}h_{b,1}^2}{h_{b,2}^2} - \frac{\sigma _{b,2}^2}{h_{b,2}^2},
\ee
where $a_1 = {{V\eta \zeta } \over {B\ln 2}} + (V + {Q_2}(t))({y_{b2}}h_{b,1}^2 + {y_{e2}}h_{e,1}^2) + (V + {Q_1}(t)){y_{e1}}h_{e,1}^2 - {{(V + {Q_2}(t))h_{b,1}^2({{V\eta \zeta } \over {B\ln 2(V + {Q_2}(t))}} + {y_{e2}}h_{e,2}^2)} \over {h_{b,2}^2}}$, $b_1 = (\sigma _{b,1}^2/h_{b,1}^2 + \sigma _{e,2}^2/h_{e,1}^2 - {{(V + {Q_1}(t))} \over {a1}} - {{(V + {Q_2}(t))} \over {a1}})$, and $b_2 = {{\sigma _{e,2}^2\sigma _{b,1}^2} \over {h_{e,1}^2h_{b,1}^2}} - {{(V + {Q_2}(t))} \over {a1}}\sigma _{b,1}^2/h_{b,1}^2 - {{(V + {Q_1}(t))} \over {a1}}\sigma_{e,2}^2/h_{e,1}^2$.

\section{Simulation Results}
\label{Simulation}
In this section, simulation results are provided to evaluate the proposed algorithm. The simulation settings are based on the works in \cite{qiot}, \cite{fzsec}. We consider the configuration with 2 UEs, which can be readily extended to a more general case. The system bandwidth for computation offloading is set as $B=1$ MHz, the time slot duration is $\tau=1$ sec, path-loss exponent is $\theta=4$, the noise variance is $\sigma_{i,j}=-60$ dBm, where $i \in \{b,e\}, j\in\{1,2\}$. The size of the arrival workload $A_n(t)$ is uniformly distributed within $[1,2]\times 10^6$ bits \cite{lyadeng}. Other parameter settings include the reference distance $d_0=1$ m, $g_0 =-40$ dB, $d_{b,1}=80$ m, $d_{b,2}=40$ m, $d_{e,1}=120$ m, $d_{e,2}=80$ m. $\kappa_n = 10^{-28}$, $P_{\max} = 2$ W, $f_{\max} = 2.15$ GHz, $C_n=737.5$ cycles/bit, the amplifier coefficient $\zeta=1$, and the control parameter $V=10^7$. The numerical results are obtained by averaging over $1000$ random channel realizations.
We consider two more cases as the  benchmark schemes to compare with our proposed algorithm. In the first benchmark scheme, marked as "Full offloading", all the tasks are offloaded to the MEC server and there is no local computation at all. The second benchmark \cite{fzsec} is marked as "Eve fully decode", in which the Eve can correctly decode other users' information. This provides a worst-case scenario for comparison. 

\begin{figure}[h]
		\vspace{-0.3cm}
	\setlength{\abovecaptionskip}{-0.2cm} 
	\setlength{\belowcaptionskip}{-1cm}
	\centering
	\includegraphics[width=2.8in]{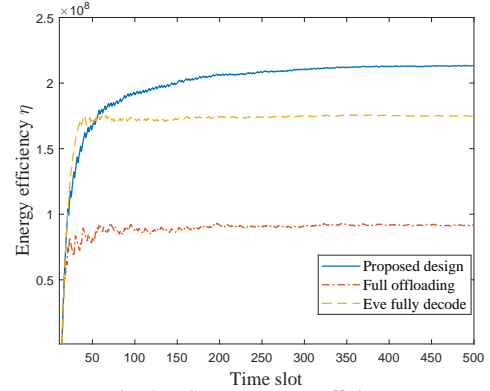}	
	\caption{System energy efficiency.\label{slotsy}}
\end{figure}
The performance of the system EE vs time is presented in Fig. \ref{slotsy}. We can see that the proposed method can achieve the highest system EE compared with the other two benchmark schemes. Furthermore, owing to the flexibility of having both offloading and local computing in the proposed scheme and in the ``Eve fully decode" scheme, the system can decide not to offload if the eavesdropper has a better channel on the offloading link while it can decide to offload if the link is secure enough. Therefore, these two schemes have a higher EE performance than the ``Full offloading" scheme, which has to offload even when the links are insecure. The system EE stabilizes for all the three schemes after $200$ time slots.

\begin{figure}[h]
		\vspace{-0.3cm}
	\setlength{\abovecaptionskip}{-0.2cm} 
	\setlength{\belowcaptionskip}{-1cm}
	\centering
	\includegraphics[width=2.8in]{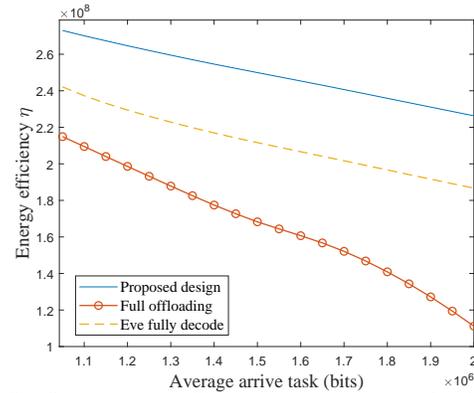}	
	\caption{System energy efficiency v.s. Average arrival task length.\label{atch}}
\end{figure}
The system EE versus the average arrival task length is presented in Fig. \ref{atch}. The proposed method achieves the highest EE. For all the three schemes, EE decrease with the increase of the arrival task length because a higher workload forces the system to increase the computing rate to maintain the low queue level. This in turn decreases the system EE. Furthermore, we notice that the performance gap between the "Full offloading" scheme and other two schemes goes up with the increase of the task length. This demonstrates that local computing is more energy efficient and secure for processing the computation tasks when the task size goes up. 

\begin{figure}[h]
	\vspace{-0.3cm}
	\setlength{\abovecaptionskip}{-0.2cm} 
	\setlength{\belowcaptionskip}{-1cm}
	\centering
	\includegraphics[width=2.8in]{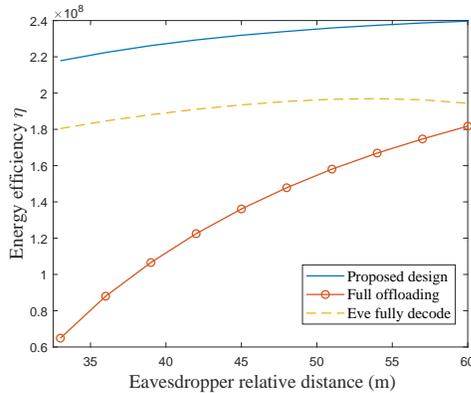}	
	\caption{System energy efficiency v.s.  eavesdropper relative distance.\label{edch}}
\end{figure}
Fig. \ref{edch} shows the system EE versus the eavesdropper location. Here the eavesdropper relative distance is defined as the distance between the eavesdropper and the UE. The proposed design achieves the best performance among all the schemes. The system EE of all the schemes goes up as the eavesdropper relative distance increases since a larger distance leads to a worse intercepting channel at the eavesdropper. Furthermore, the performance gap between the ``Full offloading" scheme and the other two schemes decreases quickly with the increase of the relative distance. This is because the secure offloading rate increases quickly when the eavesdropper moves away.  

\begin{figure}[h]
	\vspace{-0.3cm}
	\setlength{\abovecaptionskip}{-0.2cm} 
	\setlength{\belowcaptionskip}{-1cm}
	\centering
	\includegraphics[width=2.8in]{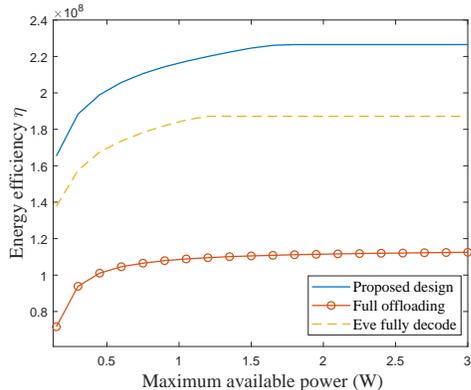}	
	\caption{System energy efficiency v.s. maximum available power $P_{\max}$.\label{pch}}
\end{figure}
The relationship between EE  and the maximum available power is illustrated in Fig. \ref{pch}. It is observed that EE increases with available power and gradually converges to a constant value. This is because that when the available power is limited, the higher computing rate and corresponding optimal EE cannot be achieved. With the power increase, EE of all the schemes keeps increasing and only stops when it achieves the highest level. After the optimal tradeoff has been reached, even there is more power available in the system, all the schemes maintain at the highest level without consuming any more power.

\section{Conclusion}
This paper aims to design a secure and energy efficient computation offloading scheme in a NOMA enabled MEC network with the presence of a malicious eavesdropper. In order to achieve a long term performance gain by considering dynamic task arrivals and fading channels, we proposed a secure task offloading and computation resource allocation scheme that aims to maximize the long-term average EE and used Lyapunov optimization framework to solve the problem. Numerical results validated the advantages of the proposed design via comparisons with two other benchmark schemes. 

\end{document}